# 3DARVisualizer: Debugging 3D Models using Augmented Reality

Srinjita Bhaduri
Department of Computer Science
Institute of Cognitive Science
University of Colorado Boulder
USA
srinjita.bhaduri@colorado.edu

Peter Gyory
ATLAS Institute
University of Colorado Boulder
USA
peter.gyroy@colorado.edu

Tamara Sumner
Institute of Cognitive Science
Department of Computer Science
University of Colorado Boulder
USA
sumner@colorado.edu

## ABSTRACT
Often neglected in traditional education, spatial thinking has proved to play a critical role in achievement in science, technology, engineering, and mathematics (STEM) education. Spatial thinking skills can be enhanced by training, life experience, and practice. One approach to train these skills is through 3D modeling (also known as Computer-Aided Design or CAD). Although 3D modeling tools have shown promising results in training and enhancing spatial thinking skills in undergraduate engineering students, when it comes to novices especially middle- and high-school students, they are not sufficient in providing rich 3D experience since the 3D models created in CAD are isolated from the actual 3D physical world. Resulting in novice students finding it difficult to create error-free 3D models that would 3D print successfully. This leads to student frustration where students are not motivated to create 3D models themselves instead, they prefer to download them from online repositories. To address this problem, researchers are focusing on integrating 3D models and displays into the physical world with the help of technologies like Augmented Reality (AR). In this demo, we present an AR application – *3DARVisualizer* that helps us explore the role of AR as a 3D model debugger including enhancing 3D modeling abilities and spatial thinking skills of middle- and high-school students.

## Tools, Skills and Materials
• Tools➡3D Printing • Tools➡Augmented Reality • Skills➡3D Modeling, Spatial Thinking

## Keywords
3D Modeling; Augmented Reality (AR); 3D Printing; Spatial Thinking skills; Middle School; High School.

## 1. BACKGROUND
Spatial thinking or "*… a mental process that involves thinking about relationships between three-dimensional (3D) objects*" [11, pp.75], plays a critical role in science, technology, engineering, and mathematics (STEM) education. Spatial thinking skills can be deepened by training, life experience, and practice [9]. One approach to train these skills is through 3D modeling [6, 15]. 3D modeling requires complex mental processes involving several steps and design decisions to make sure a feasible design is created [12]. 3D modeling and 3D printing support student learning by encouraging tinkering and making when they engage in creating and printing 3D models [4]. Given the benefits of 3D modeling, educators increasingly see it as an advantageous instructional tool in classrooms and after-school programs. However, learning 3D modeling can often be challenging, especially when it comes to middle- and high-school students because of their lack of spatial thinking skills and the complex 3D modeling software they use.

3D modeling tools often decompose the 3D designs into sequences of 2D interface operations [8]. Models created in modern 3D modeling software are not real 3D in some sense since they are isolated from the physical world. Users are not able to touch or grasp the virtual models. To address this problem, researchers are focusing on separating graphics out of conventional displays and integrating them into the physical world. Augmented Reality (AR) [3] technology provides a solution to this problem. As Azuma [4, pp: 356] states, "*AR allows the user to see the real world, with virtual objects superimposed upon or composited with the real world. Therefore, AR supplements reality, rather than completely replacing it*." AR, when integrated with CAD, is beneficial in improving students' spatial cognition abilities, supports concept development, and aids in their decision-making process along with the ability to modify designs and make refinements due to the support for viewing and 'touching' the design [13].

Modern CAD software provides a rich experience for scaling of objects, rotation, cross-sectioning, transforming, and 3D views and AR supports perspective-taking and transforming the inter-relations of objects. To take advantage of the different features of 3D modeling and AR, researchers have designed tools that enable 3D modeling within an AR context. In the early 2000s, researchers developed Construct3D [10], a 3D construction tool in AR for 3D geometry education. Users had the ability to select and interact with various geometries, customize geometric attributes, and perform several geometry operations. However, the system was expensive, and the setup was complicated, leading to restricted distribution to only a few labs so far. Other researchers have taken advantage of using AR to support creating 3D models in the real world. Do and Lee [7] have created a system called 3DARModeler that brings the conventional ways of 3D modeling to AR. Users have the ability to create complex models by assembling primitive geometries (like cubes and spheres) and grouping them to form a 3D object. The 3DARModeler allows selecting, copying, moving, pasting and deleting 3D objects in AR. Schlaug (2010) [14] on the other hand, has designed an app similar to 3DARModeler, but it uses fewer markers and is much easier to use. Black and white markers are used to augment virtual objects into the real world. Researchers like Tang et. al. (2017) [17], have designed a mobile application that assists students to learn 3D modeling skills and concepts. Using smartphones, users build models with primitive blocks in a "bottom-up" manner like "LEGO" bricks. These blocks are visualized on printed marker cards that allow users to manipulate them in the same way of manipulating real building blocks. Results have shown that the app helps users to model 3D objects in AR easily.

The tools and applications discussed, enable 3D modeling within the AR environment. Some of them have been designed to cater to specific purposes. Our app builds on these prior works and we aim to investigate how AR can be used only as a medium to support the existing 3D modeling tools. We use AR and 3D modeling in a way that allows for designing complex shapes in CAD and debugging the 3D models in *3DARVisualizer*. Thus, preserving the overall CAD experience but using AR to provide a scaffold for creating 3D designs. In this article, we elaborate on two versions of our app and studies we conducted with middle- and high-school students to test the feasibility of the app.

## 2. DEMO DESCRIPTION
In this section, we describe the techniques we followed to create *3DARVisualizer*.

### 2.1 Description of the Product/Project
The motivation behind creating *3DARVisualizer* was the prior experience of the first author facilitating 3D printing curriculum with K-12 youth. Youth participating in various 3D printing and 3D modeling curriculum come from different backgrounds with a range of skillsets. Some are knowledgeable in using 3D modeling software while some are novices or first-time 3D modelers. It becomes a challenge for facilitators to keep them engaged since they are often not able to follow along with the facilitator due to the different learnability and usability challenges they face [5, 8]. This often results in youth creating faulty models with various alignment issues and gaps leading to 3D print failures. Due to these failures occurring repeatedly, novices cease to create their models and instead download 3D models from online repositories, like Thingiverse [18] and 3D print them. Hence, there is no agency over the 3D models created by youth, and it results in less effective 3D modeling training and not enough support for enhancing student spatial thinking skills. To address this challenge, we developed the *3DARVisualizer* to pilot test the concept of using AR to debug 3D models for gaps and alignment issues. This version allowed for testing pre-designed 3D models and had limited features (described later). After receiving feedback on the first version we developed a more generic version where users could upload their 3D models (*.stl files) and test it. Both the versions were tested with middle- and high-school students and is described in detail in the following section. **Note:** We refer to the students as youth since they were in after-school programs.

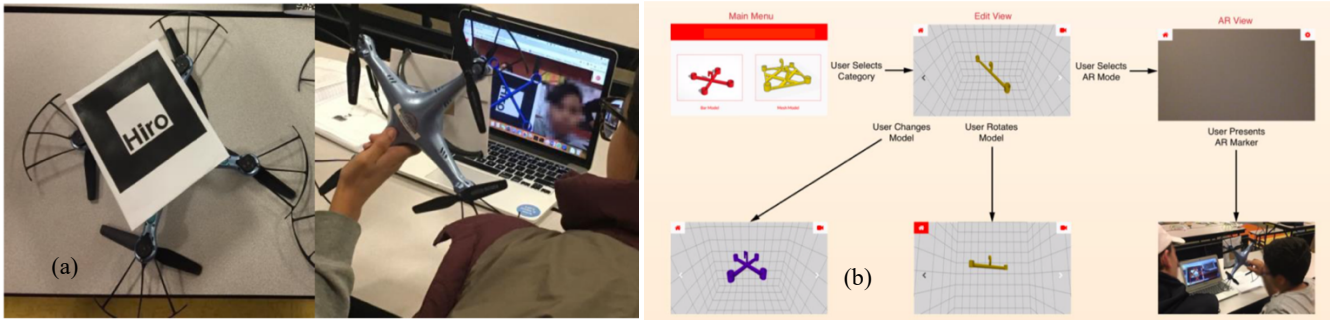

*Figure 1(a) AR marker used for training the application and youth testing a skyhook 3D model in AR with a marker attached to a UAV, (b) user flow diagram for version 1 of 3DARVisualizer*

*Version 1:*

*3DARVisualizer* was designed for a very specific after-school engineering curriculum based on Unmanned Aerial Vehicles (UAV/Drones), where participating youth must use and modify the UAVs as appropriate to conduct a range of scientific investigations, culminating in the aerial survey of a mock town suffering from a natural disaster. As part of the curriculum, youth tested 3D models of skyhooks, a hook-like attachment that fits on the four legs of UAVs. *3DARVisualizer* allowed youth to view 3D models in two different environments, 1) in a virtual environment i.e., on the laptop screens, and 2) augmented in the real-world using marker-based AR (Figure 1(a)). The reason for creating two different environments was to understand youth's ability to view the 3D models using 1) AR or 2) on a laptop screen as a sequence of 2D shapes and manipulate them from different perspectives and decide the one that would fit best with UAVs.

The core library used to render the models for this application was the open-source library Three.js [1]. To support AR functionality, we used AR.js, which is built using a Javascript port of ARtoolkit [2] as an extension of Three.js [1]. We found a web-based development approach to be ideal for this iteration of the application as it enabled ease of distribution through web browsers, rapid development cycles, and Javascript already supports many of our desired features through basic browser APIs and its vast collection of libraries. We chose a specific type of marker image (see Figure 1(a)) that supported the most precise placement of the model in the augmented context (see Figure 1(a)). The first screen (see Figure 1(b)) of the application gives the user an option of two different model categories to choose: bar and mesh. The total number of models that the user can choose from is seven (4 bar, 3 mesh) (see Figure 1(b)). For this iteration, we developed the skyhook 3D models ahead of time. When the user selects a category of model they are taken to the 'Edit View' screen, which allows them to cycle through the various models in that category and spin them around in a 3D environment. At any time while on the 'Edit View' screen, the user can select the camera button which will take them to the 'AR View' screen. The 'AR View' screen shows a video stream that comes from a connected webcam which is detected by the application. Upon detection of the marker that the program is trained to recognize, the selected 3D model is displayed on that part of the video.

***Version 2:***

With feedback from version 1, we updated the app to allow youth to upload any 3D models they create. The app was trained on a similar marker image and upon detection of the marker, the 3D model is displayed or superimposed on the marker. On opening the app in a browser, the camera is given access (see Figure 2(a)). The user has the option to upload a 3D model saved locally on their laptops or hand-held devices and is in *.stl format. After uploading, the 3D models are scaled to a specific size for the user to debug the 3D model. The app supports rotation and scaling of 3D models once they are loaded on the screen. There are buttons on the left and right sides of the screen to rotate on the x- and z-axis and the marker rotates the 3D model on the y-axis. There are zoom-in and zoom-out buttons at the bottom of the screen (see Figure 2(b)). These features allow the user to test 3D models for gaps, alignment, and overall printability. Moreover, the user can opt to test other 3D models by selecting the folder icon on the top right hand of the screen (Figure 2(b)). The core libraries used for supporting the AR functionality are similar to the first version and the *.stl file loading was implemented using Three.js STLLoader [16].

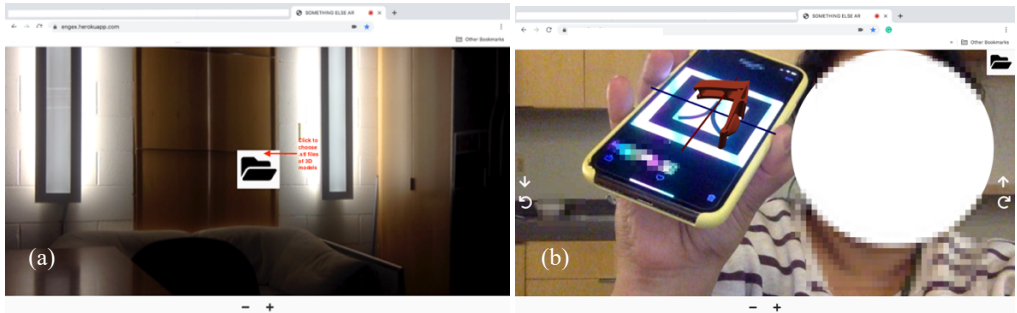

*Figure 2(a) Screen when app is given access to the camera, (b) user testing a 3D model with the AR marker image on their phone*

## 2.2 Target Audience

We tested *3DARVisualizer* in two different settings: version 1 in an after-school engineering program with 5 middle-school youth and version 2 in a 3D printing summer program with 12 high-school youth. In our future implementations, we see the app being used by teachers and facilitators of 3D printing curriculum both in formal and informal educational settings. This app aims to improve both 3D modeling and spatial thinking skills of students in middle- and high-school, especially novices.

## 3. CONCLUSION

We report on our findings from the two studies we conducted.

## 3.1 Results and Benefits

For version 1, we collected observations and screen recordings of youth using the app, we found youth had trouble with different 3D perspectives and views in the 'Edit View' screen, like the top, front, bottom, left, and right view. They did not intuitively rotate and look at the 3D models from different angles. But, when asked to use the 'AR screen' where they could visualize the 3D models of skyhooks augmented to the UAVs, youth were able to better understand the shape and design of the skyhooks. They even rotated the UAVs to look at the skyhooks from different angles and understand its fit and shape. Youth mentioned that the app helped them visualize the skyhooks "better" (Figure 1(b)). For example, using the app they were able to understand if a skyhook would fit diagonally on the UAVs legs or one side of it. This finding helped us understand the benefits of including AR with 3D modeling and we iterated on our app.

In version 2, we observed that testing the models before printing them resulted in successful prints and youth were able to test for any gaps, alignment issues, and overall printability of the models. In the surveys, 7 out of 12 youth rated *3DARVisualizer* to be 'somewhat useful' or 'very useful' in debugging their 3D models. One youth with less 3D modeling experience said, "*I was able to see many open spaces in my chess pieces in particular*" [P6] since chess pieces consist of 3D shapes stacked together. While others who had prior experience with 3D modeling did not find it very helpful (5 of the 12). We received suggestions for extensions to the debugger, such as ability to check the size, shape, and fit of the model using the app. For example, one participant said, "*It only works on a marker which isn't scaled properly, so this AR isn't as useful for checking sizes*" [P4]. **Note:** Link to video of youth testing their 3D models using version 2 of *3DARVisualizer*.

## 3.2 Lessons Learned

Keeping in mind the suggestions we collected from youth feedback, we plan on expanding version 2 of *3DARVisualizer* to include other features, like check scale of the 3D models. We are interested in testing the app on other devices and training the app with a different marker since the marker we use currently creates issues with the surrounding brightness. If there is a reflection on the marker, the 3D models do not show up. However, we did find that using 3D model debugging tools, like *3DARVisualizer*, can lower the frustration level from failed prints due to the opportunities it provides to navigate models in 3D space. It also motivates students to understand rapid prototyping opportunities that fabrication technologies offer and does have potential to enhance spatial thinking skills in middle and high school students.

### 3.3 Broader Value

Our research suggests that using Augmented Reality as a scaffold to teach 3D modeling can be valuable for promoting and maintaining interest in these digital fabrication technologies and in the process enhance spatial awareness in students. While there is work that needs to be done to generalize our findings, the current results are promising, potentially offering a new way for facilitators to use such tools in their curriculum. Moreover, we would like to modify *3DARVisualizer* and incorporate the changes that youth suggested, like easier navigation and real-life size comparison of 3D models. Finally, we plan to conduct workshops and interviews with makerspace instructors or teachers and find ways to make the AR app useful for students especially in middle- and high-school.

## 4. REQUIREMENTS

To demo *3DARVisualizer*, we will require – 1) power outlet for a laptop where the application will be demoed and 2) table for setting up a laptop and display 3D printed objects and puzzles that give a better understanding of spatial thinking skills.

## 5. BIOS

The version of *3DARVisualizer* reported in this manuscript was developed by the first and second authors. This project is part of the first author's PhD thesis and the app is being developed and iterated based on feedback the team receives. The first author will demo the app at the FabLearn Flagship conference.